\documentstyle[prl,floats,aps]{revtex}
\begin{document}
\draft
\begin{title}
{Zeros of the order parameter of $d_{x^2-y^2}$ superconducting
film in the presence of uniform current}
\end{title} 
\author{V.V. Kabanov}
\address
{\it Josef Stefan Institute 1001, Ljubljana, Slovenia}
\maketitle

\begin{abstract}
We show that additional $p$-wave component is generated in the pure $d$-wave
superconductor in the presence of the uniform current. When the current
flows in the antinodal direction the spectrum has a gap over all Fermi
surface. If the current flows in the nodal direction, the gap opens only in the
direction parallel to the current. The correction to the current due to 
the $p$-wave component of the order parameter is linear in vector
potential.  We show, that the spin-orbit coupling is responsible for
the additional component of the order parameter.  
\end{abstract} 

Recently it was proposed that classification of the order parameter in
the ferromagnetic superconductor may be performed in accordance with
corepresentations of nonunitary magnetic group\cite{fomin1,fomin2,mineev}.
Similar approach was used in Ref.\cite{kaur} to consider the effect of
broken time-reversal symmetry at $T^{*}$ on the superconducting state in
high-$T_{c}$ superconductors. Here we would like to extend this type of
analysis for the case of a $d$-wave superconductor with uniform current or/and
the external magnetic field. 


There is a common belief that the high temperature superconductors have
nontrivial order parameter transforming as $B_{1g}$ ($x^2-y^2$) representation
of $D_{4h}$ point group. When external magnetic field is applied the symmetry
of the order parameter is reduced\cite{bal1,tach}. As a result, for magnetic
field directed along the c-axis the new order parameter has the form:
$d_{x^2-y^2}+id_{xy}$. Furthermore, additional component of the order parameter
may be induced by the surface. Since any surface breaks inversion
symmetry, additional  $p$ wave component is generated near the
surface due to the surface induced spin orbit coupling\cite{edegor}.
In the following we show, that a constant superconducting current reduces 
the symmetry of the order parameter to $d_{x^2-y^2}+ip_{x}$ when the current
is directed along the x-axis. Such form of the order parameter is
consistent with the group theory analysis and may be generated by the
spin-orbit coupling. We will construct the corresponding invariant terms in
the free energy and show how $p$-wave is generated from the microscopic BCS
theory.

Since a constant superfluid current is a real vector that changes sign
under  time reversal transformation, the total $D_{4h}$ group will be reduced
to nonunitary (magnetic) group $D_{2h}(C_{2v})$. We consider for simplicity
that superfluid current is directed along the $x$-axis. Unitary subgroup
$C_{2v}$ has four elements $E,C_{2x},\sigma_{h},\sigma_{y}$ and four
representations $A_1,A_2,B_1,B_2$. In accordance with the general theory of
corepresentations\cite{erem} all 4 irreducible representations belong to the
class 'a'. This means that each irreducible representation generates one
nonequivalent corepresentation of the nonunitary group. All corepresentations
are one dimensional and are listed in the Table 1 together with the basis 
functions.  
\begin{table}[!h]\centering
\caption{Corepresentations of the nonunitrary group, when current flows in the
$x$ direction.} 
\begin{tabular}[!h]{llllllllll}
& $E$ & $C_{2x}$ & $\sigma_h$ & $\sigma_y$ & $RI$ & $R\sigma_x$ & 
$RC_{2z}$ & $RC_{2y}$ & basis functions \\ \hline 
$A_1$ & 1 & 1 & 1 & 1 & $\epsilon$ & $\epsilon$ & $\epsilon$
&  $\epsilon$ & $k_{x}^{2}-k_{y}^{2}+ia\sigma_{3}k_{y}$ \\ 
$A_2$ & 1 & 1 & -1 & -1 & $\epsilon$ & $\epsilon$ &
-$\epsilon$ &  -$\epsilon$ & 
$k_{y}k_{z}+ia\sigma_{1}k_{x}+ib\sigma_{2}k_{y}$\\ 
$B_1$ & 1 & -1 & 1 & -1 & $\epsilon$ & -$\epsilon$ &
$\epsilon$ &  -$\epsilon$ & 
$k_{x}k_{y}+ia\sigma_{3}k_{x}$\\ 
$B_2$ & 1 & -1 & -1 & 1 & $\epsilon$ & -$\epsilon$ &
-$\epsilon$ &  $\epsilon$ & 
$k_{x}k_{z}+ia\sigma_{1}k_{y}+ib\sigma_{2}k_{x}$\\ \hline
\end{tabular}
\end{table}

\noindent Here $a$ and $b$ are real numbers and $\epsilon = \exp{(2i\phi)}$ 
where $\phi$ is real. Following Ref.\cite{fomin2} we introduced factor 2 in 
the exponent. To clarify the derivation of basis functions in Table
1 we construct the basis function for the $A_{1}$ corepresentation.  It is 
easy to see that $\exp{(-i\phi)}(k_x^2-k_y^2)$ transforms as the $A_1$
representation of the unitary subgroup as well as corresponding
corepresentation of the nonunitary group. Since the group in the presence of
uniform current does not have the center of inversion, a mixture of singlet 
and triplet order parameter is possible. The general expression for the 
triplet part of the order parameter has the following form: 
\begin{equation}
{\mathbf{\Psi}}_{1} = \hat{x}f_{x}(k_{x},k_{y})+\hat{y}f_{y}(k_{x},k_{y})
+\hat{z}f_{z}(k_{x},k_{y})
\end{equation}
where $\hat{x},\hat{y},\hat{z}$ are spin unit vectors in the
${\mathbf{a,b,c}}$ axis directions respectively. The functions
$f_{x,y,z}(k_{x},k_{y})$ are odd with respect to the 
${\mathbf{k}}$ $\to$ $-{\mathbf{k}}$ transformation. 
Applying the $\sigma_{h}$ operation to $f_{x,y}(k_{x},k_{y})$ we obtain that  
$f_{x,y}(k_{x},k_{y}) = -f_{x,y}(k_{x},k_{y})$ for $A_{1}$ and $B_{1}$
representations. Therefore $f_{x,y}(k_{x},k_{y}) =0$ for $A_{1}$ and $B_{1}$
cases. Applying all operations from Table 1 to the remaining function 
$f_{z}(k_{x},k_{y})$ we obtain the following set of equations:
\begin{eqnarray}
f_{z}(k_{x},-k_{y}) = -f_{z}(k_{x},k_{y}) \nonumber \\
f_{z}(-k_{x},-k_{y})^{*} = \epsilon f_{z}(k_{x},k_{y})\nonumber \\
f_{z}(-k_{x},k_{y})^{*} = -\epsilon f_{z}(k_{x},k_{y})
\end{eqnarray}
It is easy to see that the function 
\begin{equation}
\Psi(k_{x},k_{y}) =\exp{(-i\phi)}(d_{x^{2}-y^{2}}+i\eta p_{x})
\end{equation}
where $p_{x} = \hat{z} k_{y}$ 
satisfies all the equations and transforms as the $A_1$ corepresentation of
the nonunitary group $D_{2h}(C_{2v})$. This means that the uniform current in
the $x$ direction generates the $p_x$ wave contribution to the gap
function. The gap function in that case is defined as 
$\hat{\Delta}(k_{x},k_{y})=i(d_{x^{2}-y^{2}}+i\eta
\sigma_{3}k_{y})\sigma_{2}$.  Since 
$\hat{\Delta}\hat{\Delta}^{\dag} \propto \hat{\sigma}_{0}$ the corresponding
phase is unitary and the gap in the excitation spectrum is determined by 
$|d_{x^{2}-y^{2}}|^{2}+\eta^{2}k_{y}^{2}$\cite{sigrist}. As a result, the 
gap is fully developed over all Fermi surface. 

Let us show that the term in the free energy: 
\begin{equation}
i \eta^{'} (v_x  p_x - v_y p_y) d_{x^2-y^2}^* + c.c.,
\end{equation}
is an invariant of the group $D_{4h}$. Here $\eta^{'}$ is a real constant and
$(v_x,v_y)$ are components of the superfluid velocity. Indeed, $(v_x,v_y)$ and
$(p_x,p_y)$ transform as the $E_u$ representations of the group $D_{4h}$. 
Since  the direct product $E_u \times E_u = A_{1g}+A_{2g}+B_{1g}+B_{2g}$
contains $B_{1g}$($d_{x^2-y^2}$) representation, Eq.(4) represents the true
scalar.  The presence of $i$ in Eq(4) is important since the superfluid
current is antisymmetric with respect to the time reversal symmetry.

The existence of the $p$ component of the order parameter in the presence of 
the supercurrent may be seen if we write relevant Lifshitz invariant in the
free energy \cite{minsam}:
\begin{equation}
F=F_{d}+i\eta^{''}(\psi_{d}^{*}(D_{x}p_{x}-D_{y}p_{y})-\psi_{d}
(D_{x}^{*}p_{x}^{*}-D_{y}^{*}p_{y}^{*}))+\alpha_{p}(|p_{x}|^{2}+|p_{y}|^{2})
\end{equation}
where $D_{l}=-i\nabla_{l}-2eA_{l}$ ($l=x,y,z$), ${\mathbf{A}}$ is a vector
potential, and $F_{d}$ is the free energy for the $d$-wave superconductor
without curren. Assuming that $\psi_{d}=\psi_{d0}exp{(imvx)}$ and 
$(p_{x},p_{y})=(p_{x},0)exp{(imvx)}$ we obtain a contribution to the free
energy, which is similar to Eq.(4). By minimizing Eq.(5) with respect to
$p_{x}$ we obtain the amplitude of the p-component of the order parameter: 
$p_{x}= i\eta^{''}mv\psi_{d}/\alpha_{p}$. 

When the current is flowing in the nodal direction, $B_{1g}$($x^{2}-y^{2}$)
representation of the $D_{4h}$ group is no longer $A_{1}$ corepresentation of
the reduced nonunitary group. Group theoretical analysis for this
case is presented in Table 2. It should be pointed out that this case is
similar to the case II considered in the Ref.\cite{kaur}. 
\begin{table}[h]\centering
\caption{Corepresentations of the nonunitrary group, when current flows in the
nodal direction.}
\begin{tabular}[h]{llllllllll}
& $E$ & $U_{xy}$ & $\sigma_h$ & $\sigma_{\bar{x}y}$ & $RI$ & $R\sigma_{xy}$ & 
$RC_{2z}$ & $RU_{\bar{x}y}$ & basis functions \\ \hline 
$A_1$ & 1 & 1 & 1 & 1 & $\epsilon$ & $\epsilon$ & $\epsilon$
&  $\epsilon$ & $k_{x}^{2}+k_{y}^{2}+ia\sigma_{3}(k_{x}-k_{y})$ \\ 
$A_2$ & 1 & 1 & -1 & -1 & $\epsilon$ & $\epsilon$ &
-$\epsilon$ &  -$\epsilon$ & 
$(k_{x}-k_{y})k_{z}+ia(\sigma_{1}k_{y}+\sigma_{2}k_{x})+$\\
 &  &  &  &  &  &  &  &   & 
$ib(\sigma_{1}k_{x}+\sigma_{2}k_{y})+$\\ 
$B_1$ & 1 & -1 & 1 & -1 & $\epsilon$ & -$\epsilon$ &
$\epsilon$ &  -$\epsilon$ & 
$k_{x}^{2}-k_{y}^{2}+ia\sigma_{3}(k_{x}+k_{y})$\\ 
$B_2$ & 1 & -1 & -1 & 1 & $\epsilon$ & -$\epsilon$ &
-$\epsilon$ &  $\epsilon$ & 
$(k_{x}+k_{y})k_{z}+ia(\sigma_{1}k_{x}-\sigma_{1}k_{y})$\\ 
 &  &  &  &  &  &  &  &   & 
$ib(\sigma_{1}k_{y}-\sigma_{2}k_{x})+$\\ \hline
\end{tabular}
\end{table}

\noindent Similarly to the case when the current flows in the $x$
direction, the corresponding phase is unitary: 
$\hat{\Delta}\hat{\Delta}^{\dag} \propto \hat{\sigma}_{0}$, and the gap
function has the form: $|d_{x^{2}-y^{2}}|^{2}+a^{2}(k_{x}+k_{y})^{2}$.
Contrary to the previous case, however, gap opens only in the direction
parallel to the current, and the spectrum remains gapless in the
direction perpendicular to the current\cite{foot}. 

Let us now discuss the correction to the current when the additional component
of the order parameter is generated. It is easy to see from Eq.(5) that
the additional contribution to the current is given by: 
\begin{equation}
j_{x}=j_{dx}+2ie\eta^{''}(\psi_{d}^{*}p_{x}-\psi_{d}p_{x}^{*})
\end{equation}
Substituting $p_{x}= i\eta^{''}mv\psi_{d}/\alpha_{p}$ to the last equation and
taking into account that in the quasiclassical limit $mv = -2eA$, correction to
the supercurrent has the form:
\begin{equation}
{\mathbf{j}}={\mathbf{j}}_{d}+4e^{2}\eta^{''^{2}}{\mathbf{A}}|\psi_{d}|^{2}
/\alpha_{p}
\end{equation}
At this point we should point out that usually the nonlinear London
equation is discussed in the case of the $d$-wave superconductors with the
nodes in the spectrum\cite{foot}. Nonlinear corrections to the London equation
appear due to the Doppler shift of the spectrum in the presence of the
superflow. As a result, in some regions of the Fermi surface the
excitation energy becomes negative  leading to the finite quasiparticle
current. However, as we have shown here, the generation of the $p$-wave 
component of the order parameter in the presence of the  superflow leads to
the correction to the supercurrent which is $linear$ in the vector potential. 
Opening of the $p$ wave component of the order parameter leads to the linear
in current correction to the penetration depth and could be detected
experimentally.

At the end we suggest one possible microscopic mechanism, which causes the
appearance of the $p$-component of the order parameter in the presence of
uniform current. We assume that current flows along the $x$ axis. Spin-orbit
coupling in that case could be written as
\[
H_{so}=i\gamma\sum_{k}\psi_{\mathbf{k}}^{\dag}
[{\mathbf{v}}\times\partial_{\mathbf{k}}]_{z}\sigma_{3}
\psi_{\mathbf{k}}
\]   
where $v$ is the superfluid velocity and $\gamma$ is the spin-orbit coupling
constant. Following Balatsky\cite{bal1} we can calculate the correction
to  the anomalous Green function 
$\hat{F}({\mathbf{k}},\omega)=\hat{\Delta}_{0}({\mathbf{k}})/
D({\mathbf{k}},\omega)$, where  
$D({\mathbf{k}},\omega)=\omega^{2}+\xi({\mathbf{k}})^{2}+
|\Delta_{0}({\mathbf{k}})|^{2}$, 
using perturbation theory due to the interaction $H_{so}$. The first
correction to $\hat{F}^{0}({\mathbf{k}},\omega)$ is:
\begin{eqnarray}
\delta \hat{F}({\mathbf{k}},\omega) &=& 
i\gamma v \hat{G^{0}}({\mathbf{k}},\omega)
\partial_{k_{y}}\sigma_{3} \hat{F}^{0}({\mathbf{k}},\omega) \nonumber\\
&=& i\gamma v \frac{(i\omega-\xi({\mathbf{k}}))\sigma_{3}}
{D({\mathbf{k}},\omega)^{2}}\sin{(k_{y})}i\sigma_{2} \nonumber\\
&=& -i\gamma v \frac{(-i\omega+\xi({\mathbf{k}}))\sigma_{1}}
{D({\mathbf{k}},\omega)^{2}}\sin{(k_{y})}
\end{eqnarray}
Here we take into account that 
$\hat{\Delta}_{0}({\mathbf{k}})=\Delta_{0}(\cos{(k_{x})}-\cos{(k_{y})})
i\sigma_{2}$, and 
$\hat{G^{0}}({\mathbf{k}},\omega)=(i\omega -
\xi({\mathbf{k}}))\sigma_{0}/D({\mathbf{k}},\omega)$. 
To estimate the $p$-wave correction to the order parameter we assume a
repulsive separable interaction in the following form:
$V_{y}({\mathbf{k}},{\mathbf{k}}^{'})= V_{y}\sin{(k_{y})}\sin{(k_{y}^{'})}$
As a result, the correction to $\Delta_{0}({\mathbf{k}})$ is given by:
\begin{eqnarray}
\Delta_{1}({\mathbf{k}}) &=& T\sum_{\omega,{\mathbf{k}}}
V_{y}\sin{(k_{y})}\sin{(k_{y}^{'})}\delta 
\hat{F}({\mathbf{k}}^{'},\omega) \nonumber\\
&=& -i C \gamma v \frac{\Delta_{0}}{E_{F}}N_{0}V_{y}\sin{(k_{y})}\sigma_{1}
\end{eqnarray}
Here $C$ is a real constant of the order of 1, and $N_{0}$ is the density of
states at the Fermi surface.

In conclusion, we have shown, that additional $p$ wave component appears in
the case of $d$-wave superconductor in the presence of uniform supercurrent. 
This effect leads to the additional contribution to the current linear in the
vector potential. It is shown that spin orbit coupling is responsible for
this effect. 

I am grateful to A.S. Alexandrov, A. Balatsky, D.F. Agterberg, D. Mihailovic,
T. Mertelj, and J. Demsar for useful discussions.

\end{document}